# Duality, Fundamentality, and Emergence

Elena Castellani[1] and Sebastian De Haro[2]
[1]Department of Humanities and Philosophy, University of Florence
[2]Trinity College, University of Cambridge
[2]Department of History and Philosophy of Science, University of Cambridge
[2]Vossius Center for the History of Humanities and Sciences, University of Amsterdam[1]
elena.castellani@unifi.it,    sd696@cam.ac.uk

14 June 2019

**Abstract**

Dualities offer new possibilities for relating fundamentality and emergence. In particular, as is the aim of this chapter to show, it may happen that the relations of fundamentality and emergence between dual theories are inverted. In other words, the direction of emergence typically found in these cases is opposite to the direction of emergence followed in the standard accounts: that is, while the standard emergence direction is that of decreasing fundamentality – in that there is emergence of less fundamental, high-level entities, out of more fundamental, low-level entities – in these cases of duality, on the contrary, a more fundamental entity can emerge out of a less fundamental one. In fact, this possibility can be traced back to the existence of different classical limits in quantum field theories and string theories.

---

[1] Contributed chapter for *The Foundation of Reality: Fundamentality, Space and Time,* edited by D. Glick, G. Darby, and A. Marmodoro, Oxford University Press.

# Contents

## 1. Introduction

The idea of duality has been at the centre of many important developments in theoretical physics in the last 50 years. Dualities are interesting phenomena for a variety of reasons: they are powerful tools in theory construction, they allow physicists to do calculations in regimes that are otherwise inaccessible to existing theories, and they give new insights into the physics of a problem.

Thus, from a philosophical point of view, they give interesting test-cases of various forms of empirical and physical equivalence. They also offer insights for emergence, especially the emergence of spacetime and fundamentality (see Rickles (2013), Dieks et al. (2015), De Haro (2015), Wüthrich (2019)). In this contribution, we aim to illustrate their import for some aspects of emergence and fundamentality that have not yet received much attention in the philosophical literature.

In general, the recent discussions of duality in the philosophical literature have focussed on three main questions: namely, what is the best formulation of dualities, how do dualities bear on theory individuation, and when can dual theories be said to be physically equivalent, i.e. to describe the same or different physical situations. More recently, also other aspects of dualities have started to be addressed in the philosophical literature: such as their heuristic uses in theory construction (De Haro (2018)) and their empirical consequences (Dardashti et al. (2018)).

Our motivation in bringing together duality, fundamentality, and emergence rests on two points. First, the fact that dualities can render a theory more tractable than its dual, depending on the context in which it is used—something that is not evident if one would regard a duality as a *mere* translation between two theories, like translating a text from English into French. By illustrating a notion of epistemic emergence in examples of dualities, we can better understand why physicists regard dualities as surprising and as useful for theory development: the point is precisely that dualities are not mere "automatic translation devices". Second, we wish to explicate the sense in which it is sometimes claimed that fundamentality in physics is not absolute but relative, because the relation of fundamentality can change depending on the energy or the distance scales involved (see for example Susskind (2013: p. 177), and some of the contributions in Castellani and Rickles (2017)).

Of course, a discussion of such philosophical notions as fundamentality and emergence requires us to be explicit about how they are intended. Therefore, let us specify, at this point, the sense in which we will use the notions of duality in physics, fundamentality, and emergence.

A duality in physics is, roughly speaking, a relation of formal equivalence between different theories. More specifically, it is an isomorphism between theories (for more details, see section 2).[2] In the cases we will consider, this relation will be between theories whose energy scales (or other significant parameter in the theory, such as a length scale or the strength of an interaction) are inverted, one with respect to the other: so that quantities in a high-energy regime are mapped to dual quantities in a low energy regime in the other theory. This is precisely the reason why duality, as a formal relation with significant interpretative consequences, offers new perspectives the interconnected notions of fundamentality and emergence.

Regarding 'emergence', let us start with the general notion which can be found in O'Connor and Wong (2002), where emergent entities are characterised as those that "arise out of more fundamental entities and yet are 'novel' or 'irreducible' with respect to them". Since our purposes in this paper are epistemic—we are interested in what can be derived, predicted, and sometimes also explained, by duality—we will take emergence to be a matter of *irreducibility*, as above; but we will consider *theories,* rather than entities in the theory's domain of application. Thus we take emergence, in general terms, to be the 'lack of derivability from an appropriate comparison class'.[3] This characterisation is general enough that it allows us to treat the cases we wish to deal with. Furthermore, lack of derivability may lead to novelty, so all our cases of emergence will contain novelty as well. This conception is close, for example, to Bedau's (1997) epistemic emergence, as we will discuss in section 2.2.[4]

Notice that the above characterisation of emergence involves two sets of theories. We will dub these the 'top' and 'bottom' theories, with the comparison class being the 'bottom' theory, and theories that emerge being the 'top' theory.

Turning now to 'fundamentality', we will adopt, in our use of the notion, the sense in which, in physics, bottom descriptions are usually thought to be more fundamental than the top ones, since the top theories can, at least *partly,* be derived from the bottom ones—emergence thus consisting in the fact that this derivation fails. Thus the relation between bottom and top (the 'comparison' involved in making a judgment of emergence) involves fundamentality. We will give more details on this point in section 2.2, just mentioning here some examples in order to illustrate the above sense of fundamentality.

---

[2]For a conceptual introduction to dualities, see De Haro et al. (2016). In this paper, we will take our notion of duality from De Haro (2016) and De Haro and Butterfield (2018). See also the contributions to the special issue on dualities, Castellani and Rickles (2017).
[3]As we have argued elsewhere (De Haro (2019: Section 2.1.1)), the notion of 'novelty' is in some respects more useful than derivability, because it is more general and covers more cases. Cases of lack of derivability are always cases of novelty, while the opposite is not the case. However, in the case of epistemic emergence, 'lack of derivability' suffices.
[4] An important distinction in the literature is between ontological and epistemic emergence, the distinction being in the kind of novelty or irreducibility that is considered. For example, ontological emergence is often associated to the irreducibility of laws, to the appearance of novel causal powers, and to supervenience; while epistemic emergence if often related to systemic features of complex systems that could not be predicted, or to 'lawlike generalizations within a special science that is irreducible to fundamental physical theory for conceptual reasons' (O'Connor and Wang (2002)). As we already mentioned, we are interested in epistemic emergence: which motivates our choice of 'irreducibility', as lack of derivability, in our notion of emergence above. For more on this, see section 2. For discussions of ontological emergence in the context of dualities, see De Haro (2015) and Dieks et al. (2015).

In physics, the top and bottom descriptions are often related by the variation of a continuous parameter (typically, an energy scale, length, or strength of an interaction), where the top and bottom theories are the endpoints of the variation. This variation takes us through a sequence of coarse-grainings: from the more fundamental (bottom) theory to the less fundamental (top) theory. The bottom theory is more fine-grained, i.e. it describes the domain in more detail. The top theory is more coarse-grained, it describes the domain in less detail, and so its applicability and its accuracy are reduced, relative to the bottom theory. An example of this is the relation between statistical mechanics (the bottom theory) and thermodynamics (the top theory), regardless of the question of emergence.

Another example is the distinction that physicists make between 'elementary' vs. 'composite' particles, characterising the elementary particles as more 'fundamental' because they are the classical particles that are the starting point for a quantum theory. We will join physicists in using this jargon, which indeed reflects their epistemic procedures—distinguishing between what is elementary and what is composite in a given theory is one way to make progress in physics towards finding an appropriate starting point for a new theory.

We wish to emphasise the epistemic import of the above notion of fundamentality, as well as of the notions of 'elementary' vs. 'composite' used here: they should not be understood ontologically in the sense of what is "out there in the world", since—as we will see—such a characterisation may well end up being problematic, when applied to these two notions. Indeed, as we will discuss with respect to the case studies of section 3, the distinction between what is elementary and what is composite may well end up being untenable, if it were to apply to the theory's domain of application—it is a feature of the theory's descriptive apparatus that distinguishes some characteristic patterns or behaviours, without correlating with a literal mereological distinction. Nevertheless, the distinction is of epistemic importance, since it can give a starting point for a new theoretical description.

Finally, let us note that when we talk about the 'description' given by a theory, we mean the theory's interpretation or the way it represents approximately a domain of application. However, as we said, this interpretation need not be invariant under the duality map; thus, it does not immediately follow that the interpretation should be taken literally, as part of the theory's ontology. This is indeed one of the main questions in the recent studies of dualities: whether the ontology of dual theories can be obtained regardless of dualities (in which case the duality typically does not leave the ontology invariant) or whether the theory's ontology is described only by those parts of the theory that are isomorphic under duality—what is usually called the 'common core' between the two theories that is left invariant under the duality—so that the duality should be part of one's starting point in constructing the theory's ontology. While we cannot go in detail into this discussion here—nor will our paper depend on it—let us say that our position is a cautious one, which admits a common core interpretation as the most natural interpretation of the duality in some cases, but not in others (the main cases, of sine-Gordon/Thirring and gauge-gravity dualities, discussed in this paper, do admit common-core interpretations).

Anyway, we are not specifically interested in ontological aspect here, but rather in analysing in more detail how different theoretical descriptions describe their intended domain, and in the specific advantages of particular descriptions for given domains. For example, different theoretical descriptions of the same domain may emphasise different aspects, which may even be novel with respect to each other, or incompatible with respect to one another—hence the relevance of the notion of emergence. A useful analogy may be with the different coordinate patches of a manifold: different coordinate patches describe different regions of the manifold and ascribe different local properties to it that only need be compatible at the overlaps—outside the overlaps, the properties may be very different. This is all regardless of whether a global, single-patch description exists or not: in some cases, such a global description will exist, and in others it will not.

The structure of the paper is as follows. In section 2, we discuss in some detail the notions of duality, fundamentality, and emergence. In section 3, we illustrate the interplay of these notions by means of two case-studies; in section 4, we discuss to what extent the above analysis can suggest a new way to construe the relation between fundamentality and emergence, and thereby conclude.

## 2. Duality, Fundamentality, and Emergence

As we said, in this section we discuss in turn the notions of duality (section 2.1), fundamentality (section 2.2) and emergence, of the weak, epistemic kind (section 2.3).

### 2.1. Dualities in physics

Dualities apply to a wide variety of theories, ranging from condensed matter physics to classical and quantum field theory to string theory. Our focus will be with dualities in classical and quantum field theories and string theory. First, there is the dual resonance model of the late sixties, from which early string theory originated. Successively, one of the most important developments of this idea was the generalization, proposed by Claus Montonen and David Olive in 1977, of electromagnetic duality in the framework of quantum field theory. This was later extended to the context of string theory, where dualities have also spawned recent developments in fundamental physics, offering a window into non-perturbative physics, and motivating both the M theory conjecture and gauge-gravity duality (see section 3.3).

The analogy and contrast between dualities and symmetries can be helpful before we characterise dualities more precisely. While a symmetry, in physics, is a relation within a *single* theory (typically, an automorphism of the space state and-or the set of quantities of the theory), a duality is a similar kind of relation, but between *different theories* (namely, where the notion of 'automorphism' is replaced by 'isomorphism').[5]

A bit more precisely—and our discussion will not hinge on the details—a duality is a bijective map between the states and quantities of two theoretical descriptions, such that the dynamics and the values of the quantities are preserved (for details, see De Haro (2016) and De Haro and Butterfield (2018)). Indeed, duality turns out to be a matter of isomorphic theoretical descriptions of the same physics. The papers just mentioned illustrate how a number of examples of duality in string theory and in quantum field theory indeed instantiate this notion.

A well-known example of a duality is the position-momentum duality in basic quantum mechanics. Although this is an elementary example, and one does not expect it to have the interesting properties of the dualities one finds in string theory and in quantum field theory, it is nevertheless illustrative of the general notion of duality just introduced. Namely, in basic quantum mechanics one starts with an algebra of operators, that is usually the Heisenberg algebra for position, $x$, and momenta, $p$:

$$[x, p] = i\hbar \,. \tag{1}$$

Notice that Eq. (1) underlies Heisenberg's uncertainty principle for position and momentum.

The above algebra of operators, Eq. (1), can be represented in two different ways, depending on whether one takes the position, or the momentum, to have a well-defined value: which is in agreement with Heisenberg's principle. Accordingly, the wave-function that one constructs can be either a function of the position, or a function of the momentum. The transformation that relates these two representations of the wave-functions is the Fourier transformation.[6] Leaving the problem of measurement aside: one can show that all the quantities, i.e. all the matrix elements of all the quantities constructed from $x$ and $p$, can be calculated either as functions of the position basis or as functions of the momentum. The Fourier transformation relates the two, for all the matrix elements. Thus the Fourier transformation is a duality in the above sense, albeit a very simple one.

---
[5]For the relation between dualities and symmetries, see De Haro and Butterfield (2019).

Although the position-momentum duality just discussed is not a case of weak/strong coupling duality, we can draw a useful analogy with the weak/strong coupling dualities that we will introduce in section 3. In fact, the Fourier transformation has the property of "inverting the uncertainties" of the states, since it maps a well-localised wave-function to an ill-localised one. Namely, take a wave packet with spread $\Delta x$. This means that the position of the particle is known with uncertainty $\Delta x$. Let us also assume that the particle is well-localized, i.e. that $\Delta x$ is very small, compared to a relevant length scale. Then it follows from the Fourier transformation of the wave-function (alternatively, it follows from Heisenberg's uncertainty principle) that the uncertainty in the particle's momentum will at least be of order $\hbar/2\Delta x$, which is very large if $\Delta x$ is small. Thus the Fourier transformation maps a localised particle to a delocalised one.

Informally speaking, we can say that 'having a well-defined position' is a typical particle property, while 'having a well-defined momentum' is a wave-like property.[7] This can also be seen as a *duality between particle and wave-like properties*,[8] and it traditionally goes under the name of 'wave-particle duality'.

Just to give an idea, the analogy between the situation just described and the weak/strong coupling duality that we will discuss below goes as follows. In the more sophisticated cases of dualities in string theory and quantum field theory, when one theory is weakly coupled, so that the classical approximation holds good (as in the case of the well-localised particle), the other theory is strongly coupled, i.e. it is highly interacting and quantum (as in the case where the momentum is ill-defined, so that the description in terms of a single momentum value is not valid).

## 2.2. Emergence and fundamentality

In Section 1, we characterised epistemic emergence as the 'lack of derivability from an appropriate comparison class', where this lack of derivability implies that there is *novelty* in the theoretical description. Thus the novelty is, in this case, not to be found in the world but in the way the theory describes the world.

At this point, a further refinement of this 'lack of derivability' will help in clarifying our case studies: for lack of derivability can be a matter of principle—an intrinsic limitation of a theory (e.g. the theory is not general or precise enough that the emergent theory can be derived from it), or it can be a matter of practice. The latter option—lack of derivability in practice—is considered, for example, by Bedau (1997), who proposes a weak notion of emergence that allows derivation, but only if it is *by simulation*. More precisely, Bedau defines a macrostate to be weakly emergent from the microscopic dynamics iff it can be derived but only by simulation. Clearly, this lack of derivability is not one of principle; rather, it is a practical one, since a way to derive the behaviour *does* exist: that is, simulation. But, according to Bedau, "the algorithmic effort for determining the system's behaviour is roughly proportional to how far into the future the system's behaviour is derived" (p. 378). Thus, despite the in-principle availability of derivation, such a derivation is unreachable in practice for epistemic agents with limited resources (such as limited information about the initial conditions, computing power, etc.).

---

[6] A Fourier transformation is a mathematical technique widely used for waves, e.g. to decompose sound waves (which are described as the vibrations of air in space, i.e. the oscillations are functions of the position in space) into elementary frequencies (so that the oscillation function is now not a function of the position, but of the frequencies or, equivalently, of the momenta: see also footnote 6). The Fourier transformation relates the two descriptions.

[7] A wave is typically characterised by its wavelength, $\lambda$. But by de Broglie's relation between the momentum and the wavelength of the particle, viz. $\lambda = h/p$, a wave can also be characterised by its momentum. This is the reason why we say that dependence on the momentum is a typical wave-like property since, against the background of de Broglie's relation, it is also dependence on the wavelength.

[8] We are being informal here, since the aim of the example is only to illustrate a duality that is analogous to the case weak/strong coupling duality. The notion of wave-particle duality of course depends not only on Heisenberg's uncertainty principle, but also on whether the behaviour is "particle-like" or "wave-like", on detection with a measurement apparatus.

Although in this paper we are not concerned with simulation, the above example does illustrate well the distinction between weak and strong epistemic emergence that we are interested in (in this respect, see also Guay and Sartenaer (2016)). Strong epistemic emergence is the lack of derivability in principle—the theory simply lacks the resources to derive whatever is emergent from it. Weak epistemic emergence is the lack of derivability in practice—some derivations may be available, but they are difficult to carry out within the theory's methods or resources, so that the situation is, in practice, as if one was dealing with strong emergence.

We agree with much of the philosophical literature in thinking that epistemic emergence is a genuine form of emergence. In fact, we agree with Bedau (1997) that weak emergence is not to be dismissed as merely "subjective" or as referring only to human factors, since it stems from the complexity of the theoretical description, which is an objective feature of the theory. In particular, we assume that epistemic emergence has to do with the limited applicability of a theoretical description, given its own techniques and the description of the world that it gives.

Indeed, the cases of weak epistemic emergence that we will consider here will be such that, once a theoretical description ceases to be applicable because it loses its practical predictive power, a *novel description* emerges through duality that does have that predictive power and applicability. More precisely, the fact that these cases of emergence are weak depends on the assumption that we have *exact dualities* between the theoretical descriptions: then, the limitations are indeed practical rather than a matter of principle, stemming from the fact that the specific theoretical description can only be (easily) used in a limited domain. But since the duality is exact, the dual description could in principle be derived by using the duality, and then the two theories are equivalent when the full domain is considered: it is only that it is very difficult to do so in practice. Our examples from section 3.2.3 are of this type.

However, there are cases in which a proof that the duality is exact is not currently known, and even the theoretical descriptions are only approximations covering their own limited domain of phenomena, so that the two descriptions could be only approximately dual. In this case, if it turns out that the duality is approximate—that is, if it turns out that the two descriptions are not exact duals of each other, and there is no better description that renders them dual—then epistemic emergence is strong, since derivability ultimately fails. Thus a verdict whether duality is weak or strong depends on delivering a proof of the duality, and on having theoretical descriptions that are not approximations— a situation that is rarely met in physics.[9] In fact, we proceed on the plausible assumption that these dualities will one day be proven, so that epistemic emergence is weak.

So far we have talked about emergence of theoretical descriptions through duality. This indeed covers all the cases we will discuss (see sections 3.2.3, 3.2.4, and 3.3). However, as we will see, the process of finding a new, dual, description often begins with identifying new, emergent quantities, so that a problem that looked intractable in the original description, becomes the starting point for the new description, where it is tractable. Typically, these new quantities represent a new class of particles;

---

[9] For cases with a proof of duality, see De Haro and Butterfield (2018). The examples in section 3.3 are cases of dualities the existence of which is well-supported by the evidence, but which have not been proven. For a discussion of the evidence, see De Haro et al. (2017).

however, recall that we are not committed to the ontological status of these particles.

We now turn to discuss fundamentality. It is useful to keep in mind that weak emergence relates (what in physics jargon is usually called) the top and bottom theories or entities, where the *bottom* theory or entity is regarded as more fundamental, because (at least in principle) the top theory or entity can be derived from it. This one-way implication—the top theory can be derived from the bottom theory, but not the other way around—means that the bottom theory is more general than the top theory, and covers more cases, so that the bottom theory can make at least some predictions that in practice cannot be made using the top theory. There are various mechanisms for this, of which we here mention three that we will encounter in the examples:

(1) A well-known example is *coarse-graining,* where the bottom theory is more fine-grained, so that it has describes the domain of application in more detail. The top theory is more coarse-grained, it describes the domain in less detail, and so its applicability and its accuracy are reduced, relative to the bottom theory. A well-known example is the relation between statistical mechanics and thermodynamics, where the former describes the fine-grained micro-dynamics, while the latter can be derived from it (at least in important classes of examples) by using appropriate bridge laws (and this possibility to derive one from the other is of course regardless of possible emergence).
(2) Another example is provided by weak/strong dualities in the framework of perturbative quantum field theory. The perturbative Feynman diagram techniques of quantum field theory do not work at strong coupling, and there are no general and systematic methods for dealing with strong coupling situations in quantum field theory. This is why the presence of a *weak/strong coupling* duality—which allows us to transform the strong-coupling problem into a weak-coupling problem—is a significant theoretical advantage, and forms the starting point of a new theoretical description that is more fundamental in that regime, in the sense we have defined in section 2.2—it is a theoretical description with predictive power, while the original theoretical description has no predictive power. But this notion of fundamentality is *relative*, as we will see, i.e. it depends on the domain to which the theory is applied.
(3) In some cases, rather than varying a continuous parameter (or in combination with it) the duality involves a (not necessarily invertible) *change of variables,* whose definition can contain an infinite number of terms. Again, such infinite redefinitions allow us to change a theoretical description that is practically intractable into a tractable one.

**2.3.    Three options for emergence**

In (1)-(3) in the previous section, we considered changing either a single parameter (the scale, or the coupling, in (1)-(2)), or variables (3). But these options can be combined. In that case, it is not *a priori* clear what emerges from what, and so what is the fundamentality relation. Our discussion of (1), for example, assumed the direction of coarse-graining as the emergent one, and our discussion of (2) assumed that, in the variation of the coupling strength, the theories that we find at the endpoints of the variation are the relevant bottom and top theories. But what happens if we combine options (1) and (2), or combine options (2) and (3), etc., so that we have "different directions"?

We distinguish three different options for combining (1). (2) and (3):

Option (i). This is the option usually endorsed in the philosophy of science literature in connection with reduction: a finer-grained theoretical description is viewed as more fundamental than a coarser-grained theoretical description. This is the sense according to which, for example, particle physics is considered a more fundamental description (because it is more "fine-grained") of the physical world than condensed matter physics, condensed matter physics more fundamental than chemistry, chemistry more fundamental than biology, and so on.

In particle physics, in particular, the physical description and its degree of fundamentality is interpreted as related to a physical scale. Accordingly, the fact that at physical scale ranges we can have remarkably different physics has found an explicit realization in the so-called effective field theory approach. In general terms, an "effective field theory" (EFT) is a theory which "effectively" captures what is physically relevant in a given domain. More precisely, it is a convenient, appropriate description of the relevant physics in a given region of the parameter space of the physical world.[10]

A key point in the EFT approach is the separation of the physics at the chosen energy scale from the physics at much higher energies: an EFT describes the physics relevant at a given regime and this low-energy description is largely independent of the high-energy theory. In this sense one can say that the low-energy theory is emergent with respect to the high-energy one.

In fact, both the decoupling between what happens at a high-energies and what happens at a low-energies, and the emergence of new properties and behaviours at different energy ranges, were used by the Nobel Prize P. W. Anderson, in his 1972 seminal article entitled "More Is Different", for arguing against the view that high-energy physics is more fundamental than condensed matter physics (i.e. arguing against (i)).

<u>Option (ii).</u> This is the option endorsed by those who argue in the following way (with respect to the direction of emergence and its use in defining fundamentality): because of the facts of emergence and decoupling, unlike option (i), fundamentality is not to be based on the position on the energy scale, but it is to be discussed at the same scale, i.e. each and every scale, considered on its own.

<u>Option (iii).</u> This option is a bit unusual, in the sense that according to it emergence and fundamentality appear as disconnected. As we will see, it concerns cases in which what emerges in the dual description plays a fundamental role in the new dual description. To illustrate how this is possible, we will discuss two case studies, both involving weak/strong coupling duality. In particular, we will show how this kind of duality can provide interesting examples of weak epistemic emergence.

## 3. Two Case Studies

In this section, we introduce the two case studies that we will use to analyse the bearing of duality on fundamentality and emergence: namely, "generalized electric-magnetic duality" (section 3.2) and "gauge-gravity duality" (section 3.3). Since these case studies are examples of weak/strong coupling duality, we will start with briefly introducing, in section 3.1, this kind of duality in the framework of perturbation theory in quantum field theory.

Let us note, at this point, that in addition to the emergence of theories, we will also consider emergence of "effective entities": these are not the entities in the theory's domain of application, but rather variables of a theoretical description that are interpreted as effective entities that appear in the theory's domain of application when a certain *approximation* is made. The idea is that these effective entities are features of a theoretical description, with various particle-like or object-like behaviours. Thus, in the rest of the paper, our 'effective entities' should be understood in this way—they are useful features of a theoretical description with various particle-like or field-like properties.

Thus like our use of 'effective entity', which is epistemic, also our use of names such as 'particle', 'elementary', and 'composite' is epistemic. Thus our use of these words does not commit us to their existence as entities in the theory's domain of application. Rather, it reflects the effective entity's possession of particle-like properties, etc., as described by a specific theory.

---

[10] There is a lively philosophical discussion on EFTs: see for example Georgi (1997, p. 88). On the concept of a parameter space, see section 3.1 below.

## 3.1. Weak/strong coupling duality and perturbation theory

Weak/strong coupling duality has become a basic ingredient in fundamental physics, especially since the 1990s. In general terms, it is a duality such that the weak coupling regime of one theory is mapped to the strong coupling regime of the other theory. The special interest in this form of duality stems from the fact that it is seen as a new tool for getting information on physical quantities in the case of large values of the coupling constant, where the usual perturbative methods fail,[11] by exploiting the results obtained in the weak coupling regime of the dual description.

Let us unpack some of the notions used above, especially: 'couplings' (or coupling parameters), and 'perturbation theory'. A coupling is, roughly, a parameter characterising the strength of a force. Thus Newton's constant, $G_N$, is the coupling parameter of the gravitational force, and the spring constant, $k$, is the parameter characterising the strength of Hooke's law, viz. the coupling of the spring force. In Maxwell's theory, the electric charge, $e$, plays the role of the coupling.

In this respect, we will also consider another important parameter, Planck's constant $\hbar$;[12] it is the dimensionful parameter that typically indicates the importance of quantum effects (in other words, quantum effects are large or small compared to this parameter). Although Planck's constant is strictly speaking not a coupling constant in the way just described (it does not characterise the strength of a force, but rather the importance of quantum effects), we will see that it plays a similar role as the coupling constants do.

Let us write the coupling constant as $g$, for whatever force is present in the problem. Since the coupling constant characterises the strength of the force, an expansion of the physical quantities around the point $g = 0$ is an expansion subject to the assumption that the force is weak, and so that the interactions are small:

$$Q(g) = Q(0) + g\, Q_1 + g^2 Q_2 + \cdots, \qquad (1)$$

where $Q(g)$ is the quantity of interest, as a function of the coupling. The above expansion is called the 'perturbative expansion' of the theory, i.e. it is an asymptotic expansion for small interactions, or weak coupling.[13] In quantum field theories, where the interactions are of a quantum mechanical nature from the start, the above expansion turns out to *coincide* with the expansion in $\hbar$, as we will discuss in section 3.1.3. So, the first term is the classical contribution, and the sub-leading terms are quantum corrections.[14]

An important ingredient of quantum field theories is the so-called 'flow of the couplings'. Namely, unlike ordinary quantum mechanics where the coupling $g$ is a constant, the coupling in quantum field theory is a function of the momentum, $k$, i.e. $g = g(k)$, where $k$ is like energy. This has to do with the effects of *renormalization*, namely the basic fact that—due to the infinite number of particles that are assumed to be present in quantum field theory—the self-interactions and mutual interactions of fields give rise to new terms that have to be taken into account in the interactions of the theory.[15] In fact, the coupling constant $g(k)$ satisfies an equation, the renormalization group equation, which fixes the dependence of the coupling on $k$. This equation describes the 'flow' of the coupling constants (if there are more than one) in their parameter space. We will get back to this notion in section 3.3. We now turn to illustrating our two case studies.

---

[11] "Failure of perturbative methods" here means that the expansion in section 3.1 (below) does not converge, because $g$ is not small (in a weaker sense, it means that one has to take into account an infinite number of terms in this expansion, which in practice is often impossible to do). This makes dualities particularly interesting and useful in the context of quantum field theory and string theory, since we usually know only the perturbative part of a theory, that is its 'weak coupling' regime. Dualities thus can be used to relate what is still unknown to what is known.

[12] The constant $\hbar = h/2\pi$, where $h$ is Planck's constant, is called the *reduced* Planck constant. For simplicity, we will continue to call it Planck's constant.

[13] We will now not enter into the details of whether this expansion converges. This is obviously an important issue. However, in theories with dualities it is usually a good assumption (modulo technicalities), because the duality ensures that the regimes of both small and large $g$ are under control. In those cases, the difficulty will be not the convergence, but the fact that one needs to take into account an infinite number of terms (see footnote 13).

[14] This coincides with the celebrated *Feynman diagram expansion*, which may be familiar to some readers.

### 3.2. Generalized electric-magnetic duality

Electric-magnetic duality (EM duality) represents the first form of duality to be explicitly applied in twentieth century fundamental physics. The idea that there is a substantial symmetry between electricity and magnetism is an old one, dating back to the 19$^{th}$ century where it played a role in Faraday's discovery of electromagnetic induction and was first made more precise with Maxwell's formulation of his famous equations regulating the behaviour of electric and magnetic fields. In its contemporary form, its origin and first developments are due to P. A.M. Dirac's famous paper (1931 and 1948) on his "theory of magnetic poles". In fact, the very idea of weak/strong duality stems from Dirac's seminal work and its successive generalizations in the context of field and string theory.

From the viewpoint of the issue at stake here—namely the significance of duality to the discussion of fundamentality and emergence—EM duality in its generalised form is particularly interesting because of the following novel feature: that is, the fact that the weak/strong coupling nature of the duality manifests itself in the fact that under EM duality it often happens that what is viewed as 'elementary' in one description gets mapped to what is viewed as 'composite' in the dual. At first sight, this interchange between what is 'fundamental' and what is 'composite' could be taken to suggest an ontological, relative notion of fundamentality. But this reading is too quick. Actually, what this case seems to best suggest is a form of epistemic relative fundamentality or "representational fundamentality" (as argued in Castellani, 2017).

In what follows, we will enter into some details of the generalized EM duality case study, in order to identify those specific features that illustrate the option (iii) mentioned in section 2.3, in the relationship between duality, emergence and fundamentality. We will structure this brief overview of the main features of the EM duality according to its actual historical development. In 3.2.1, we discuss the classical formulation of EM duality in the context of Mawxell's electromagnetic theory, turning in 3.2.2 to its extension to the quantum context with Dirac's "Theory of Magnetic Poles". Then, in 3.2.3, we discuss the generalization of EM duality within the framework of quantum field theory.

#### 3.2.1. Electric-magnetic duality in classical electromagnetism

In Maxwell's theory, there is an evident similarity in the role of electric and magnetic fields. This similarity is complete in the absence of source terms (electric charges and currents), and this is mathematically expressed by the fact that Maxwell's equations do not change in form when the roles of the electric field E and the magnetic field B are exchanged in the following way:

$$D: \quad E \to B, \quad B \to -E$$

The transformation $D$ is a *duality transformation,* which leaves Maxwell's equations invariant.[16]

When electric source terms are present, however, the Maxwell equations are no longer invariant under the duality transformation, $D$. In order to restore the duality of the theory in the presence of source terms, one needs to postulate the existence of magnetic charges beside electric charges and, accordingly, to modify Maxwell's equations. In their new form, these equations are then invariant under the duality transformation $D'$, which at the same time exchanges the roles of the electric and magnetic fields, and of the electric and magnetic sources, as follows:

$$D': \quad E \to B, \quad\quad B \to -E \quad\quad (2)$$
$$(e, j_e) \to (g, j_g), \quad\quad (g, j_g) \to (-e, -j_e)$$

---

[15] For a philosophical review, see Butterfield and Bouatta (2015); for a brief discussion, close to our second case study where we will discuss renormalization, see Dieks et al. (2015: p. 207).
[16] Since the transformation is on the same theory, this is a case of self-duality, i.e. a symmetry.

Here, $(e, j_e)$ represents the electric charge and electric current, and $(g, j_g)$ the magnetic charge and magnetic current.[17]

There is a problem, however: isolated magnetic charges, i.e. the so-called magnetic monopoles, have never been observed. Breaking a magnet bar in two parts, one obtains two smaller magnets but never an isolated North pole or an isolated South pole. Assuming, nevertheless, the existence of magnetic charges in order to save the duality between electricity and magnetism, leaves this question to be addressed. In fact, the extension of EM duality to the quantum context, as we will see below, allowed Dirac to give the following answer: isolated magnetic poles had never been observed because an enormous amount of energy is needed to produce a particle with a single magnetic pole.

### 3.2.2. Extension to the quantum context

The extension of EM duality to the quantum context was carried out by Dirac in the two papers (1931, 1948) in which he developed his theory of magnetic monopoles. In these papers, Dirac proved that it is possible for a magnetic charge, $g$, to occur in the presence of an electric charge, $e$, without disturbing the consistency of the coupling of electromagnetism to quantum mechanics.[18] As Dirac proved, the condition for this to be possible, known as *Dirac's quantization condition,* is the following one:

$$eg = 2\pi n \hbar c \quad\quad n = 0, \pm 1, \pm 2, \ldots \quad\quad (3)$$

where $c$ is the speed of light. Dirac's condition thus established the existence of an inverse relation between electric and magnetic charge values, with many relevant consequences.[19] In particular, from the viewpoint of interest here, this condition provided the basis for the idea of weak/strong coupling duality. Indeed, by combining Dirac's condition with the fact that EM duality interchanges the roles of electric and magnetic charges, as above (i.e. combining Eqs. (2) and (3)), we obtain the following inverse relations:

$$e \to g = \frac{2\pi n \hbar c}{e}$$

$$g \to -e = -\frac{2\pi n \hbar c}{g}$$

This means that, if the charge $e$ is small (i.e. weak coupling), the dual charge $g$ is strong (strong coupling), and vice versa: in other words, in a quantum context EM duality relates weak and strong coupling. However, it is only with the generalization of EM duality to the framework of quantum field theory that the idea of weak/strong coupling duality started to acquire its modern meaning and fruitfulness. Thus, we now turn to this decisive step in the history of EM duality, with a particular focus on the related interchanging role of 'elementary' and 'composite' between the entities in the dual descriptions.

---

[17] For more detail on this and the next subsection, we refer the reader to Castellani (2010, 2017).
[18] Turning from the classical to the quantum formulation of electromagnetic theory with magnetic sources posed a consistency problem: the electromagnetic vector potential A, playing a central role in coupling electromagnetism to quantum mechanics, is introduced in standard electromagnetism by taking advantage of the absence of magnetic source terms.
[19] First, it provided an explanation of why isolated magnetic poles had never been observed. Second, it explained the quantization of the electric charge: the mere existence of a magnetic charge, $g$, somewhere in the universe would have implied the quantization of electric charge, since any electric charge should occur in integer multiples of the unit.

### 3.2.3. Sine-Gordon/Thirring duality and Montonen-Olive conjecture

Historically, the seminal contribution for the generalization of EM duality to the quantum field theories of particle physics was the 1977 work by Montonen and Olive, entitled "Magnetic monopoles as gauge particles?", where they formulated their celebrated EM duality conjecture: that is, in their own words, the conjecture that "there should be two 'dual equivalent' field formulations of the same theory in which electric (Noether) and magnetic (topological) quantum numbers exchange roles" (p. 117).

In order to understand the physical implications of this conjecture, let us take a step back and mention a previous result: namely, the duality between the so-called *sine-Gordon theory* and massive *Thirring model*.[20] This duality, which was firmly established by works of S. Coleman and S. Mandelstam in the mid-1970s, originated from pioneering contributions by T.H.R. Skyrme betwen the end of the 1950s and the beginning of the 1960s. Let us mention two things in particular: (a) his pioneering idea that a soliton could be interpreted as a quantum particle,[21] and that a dual correspondence could be established between this sort of particle—which is extended, and therefore not considered as elementary—and the familiar elementary particles of quantum field theory; (b) his conjecture that the nucleons (spin 1/2 fermionic states) could emerge as the soliton states of a purely bosonic field theory.

Skyrme's conjecture was confirmed in 1975 by Coleman and Mandelstam's work proving the dual equivalence of the sine-Gordon and massive Thirring models in general terms. From the viewpoint of this paper, we will focus on the following results:

(a) The equivalence was proven to be a weak/strong coupling duality: the weak coupling regime of the sine-Gordon fields corresponds to the strong coupling regime of the massive Thirring model, and *vice-versa*.
(b) This duality implies, in particular, a precise correspondence between the soliton states of the quantized sine-Gordon theory and the elementary particle states of the dual massive Thirring model.

In other words: by means of the weak/strong coupling duality, the sine-Gordon quantum soliton was proven to be a particle (the "elementary" fermion of the massive Thirring model) in the usual sense of the concept in particle physics.[22] Thus, in the full quantum theory, particles could appear as solitons or as elementary particles, depending on the way the theory was formulated (whether as the quantum sine-Gordon model or as the massive Thirring model): their status was equivalent. Coleman (1975, p. 2096) famously commented on this fact in terms of a situation of democracy among the particles: "Thus, I am led to conjecture a form of duality, or nuclear democracy in the sense of Chew, for this two-dimensional theory."[23]

---

[20] These are two field theories in one space and one time dimension, describing, respectively, a massless scalar field $\psi$ (with interaction density proportional to $\cos\beta\psi$) and a massive self-coupled fermionic field. See Castellani (2017, section 2.2.1).
[21] Solitons are extended solutions of classical non-linear field equations, so called by Zabusky and Kruskal (1965) to indicate humps of energy propagating and interacting without distortion. They were first discovered in nineteenth century hydrodynamics in the form of 'solitary water waves', whence their name.
[22] That is, structureless particles arising from the quantization of the wave-like excitations of the fields.
[23] On the idea of nuclear democracy in Chew's S-Matrix approach in the 1960s, according to which no hadron was more fundamental than the other, see in particular Cushing (1990). Comments on this can be found in Castellani (2017, section 3.2).

As we already mentioned in 3.2, in discussing the present case study we will take the notion of fundamentality to be based on the 'elementary vs. composite' distinction: a particle in a given theoretical description is considered to be more fundamental if it is elementary, and less fundamental if it is composite (or extended).

The exact equivalence between the two theories, i.e. result (b) above, is worth stressing. For it means that the fermionic state of the Thirring model *is already there in the sine-Gordon theory*, and vice-versa: a bosonic state of the sine-Gordon theory is already there in the massive Thirring model.[24] In this sense, there is no ontological emergence, because the two theories describe exactly the same states, quantities, and dynamics (though, as we will argue later, there is epistemic emergence).

Sine-Gordon/Thirring duality was the first explicit example of a weak/strong duality with a corresponding dual interchange of elementary particles and solitonic particles in the framework of a quantum field theory. It was therefore natural to try to extend these ideas to the more realistic case of a physical space-time of three space and one time dimensions. This was proposed by Montonen and Olive in their 1977 work, in terms of a generalization of Dirac's EM duality in the context of a unified quantum field theory of weak and electromagnetic interactions.

Just as for Dirac's theory, the duality considered by Montonen and Olive is a case of self-duality: the very same theory has two equivalent dual descriptions. What is of particular interest, here, is the kind of situation that this generalized EM duality represents: a quantum field theory describing both particles with "electric" charge $e$, and particles with "magnetic" charge $g$,[25] which can have two different classical limits, depending on which coupling (charge)—$e$ or $g$—is kept fixed while taking the classical limit $\hbar \to 0$.[26] Accordingly, there are two possible scenarios, corresponding to the two dual descriptions of the same quantum theory:

(1) If the "magnetic" coupling $g$ is kept fixed, then, from Eq. (3), the classical limit, $\hbar \to 0$, corresponds to weak electric coupling, viz. $e \to 0$: in this case, the electrically charged particles play the role of elementary particles, and the magnetically charged particles of solitons.
(2) If the "electric" coupling $e$ is kept fixed, then, from Eq. (3), the classical limit, $\hbar \to 0$, corresponds to weak magnetic coupling, viz. $g \to 0$: in this case, the magnetically charged particles play the role of elementary particles, and the electrically charged particles of solitons.

The particles, whether electrically or magnetically charged, are all present in the complete quantum theory. In this sense, they all are equally "fundamental", from an ontological point of view. What the duality implies, however, has rather to do with their different modes of appearance when considering the different classical limits of the quantum theory (the dual perspectives). They interchangeably play the role of "elementary" (i.e. "fundamental") or "solitonic" particles, depending on the perspective under which the theory is considered.

At this point, what conclusions can be drawn with respect to fundamentality and emergence in the light of the cases of weak/strong duality in QFT just illustrated? As seen, these are both cases in which a weak/strong coupling duality is accompanied by an interchange of the 'elementarity' and 'compositeness' of the particles in the quantum theory. In short, we can make the following considerations, at this stage:

(A) In the case of sine-Gordon/massive Thirring model duality, we have two different quantum theories—a bosonic field theory vs. a massive fermionic theory—which are related by a weak/strong coupling duality, such that an elementary particle in one theory becomes a soliton state in the other.

---

[24] The fermionic state is non-perturbative (i.e. not visible at weak coupling) in the sine-Gordon theory; as is the bosonic state not visible in the weakly coupled Thirring model. But the states are there nevertheless. For more details, see De Haro and Butterfield (2018: section 5.2.2).
[25] See for example Sen (1999, Section 2); Polchinski (2017, p. 7). Electric and magnetic are here to be intended in a generalized sense. For a more detailed treatment, we refer to Castellani (2017, section 2.2.3).
[26] Planck's constant $\hbar$ is of course a dimensionful constant of nature, and we cannot change its value. What we have in mind here is that we consider a sequence of semi-classical solutions of the theory, with successively larger values of the typical action in the solution (in comparison with $\hbar$) while we keep the couplings (including $\hbar$) fixed. Taking the $\hbar \to 0$ thus involves comparing different physical systems. This is the case also if one takes e.g. $e/\sqrt{\hbar}$ as one's expansion parameter. Also, notice that in the quantum field theory literature, $e$ is not measured in Coulombs, because it has been divided by the square root of the vacuum permittivity. This is the reason why $e/\sqrt{\hbar}$ can be taken to be a dimensionless parameter, and electric and magnetic charges can be related through Eq. (3).

   Regarding fundamentality: as said above, the effective entities of the two theories are ontologically equally fundamental, since all the states and operators (for both particles and solitons) are already there in the two theories. Therefore, there is no ontological emergence, as already discussed. However, we can also look at the different roles that the bosonic and fermionic particles play in the two dual *descriptions*: in one description, the bosonic particles are the elementary particles, while the fermions only emerge as solitons in the high-energy limit. In the dual description, it is the reverse. In other words: being elementary or being a soliton is a matter of the convenience of the description, i.e. it depends on the specific fields one is working with, and the relation between the two pictures is like a (admittedly, very complicated) change of variables. Thus, this kind of emergence is only weakly epistemic. If we take the elementary particles to be more fundamental than the composite ones in a given description, then the notion of fundamentality is relative to that description, and—like emergence—fundamentality in this example is relative, from an epistemic point of view.

(B) In the case of generalized EM duality (Montonen-Olive duality), the difference with the previous case is that emergence takes place within the same theoretical context, since it is a case of self-duality (i.e. the duality map does not take us out of the theory). Nevertheless, we can reach the same conclusions as in (A). That is, in one description the electric particles are elementary (and the magnetic particles are then solitons or "composite"), while in the other description it is the opposite.[27]

### 3.3. Gauge-gravity duality

Around 1995, the discovery of string dualities and of D-branes (which are extended, non-perturbative objects in string theory) motivated the idea of the existence of a relation between gravity theories and gauge theories. The microscopic entropy counting of Strominger and Vafa (1996) for extremal black holes, which is seen as one of the successes of string theory, vindicated this relation between gauge theory and gravity: since the entropy of a black hole (the gravitational object *par excellence*) is calculated by counting microstates in an associated gauge theory. In 1997, Maldacena took this relationship a step further, by relating string theory in anti-de Sitter space, or AdS (i.e. a space with a negative cosmological constant)[28] to a quantum field theory (QFT) which is scale invariant, i.e. a gauge theory.[29] This is called 'gauge-gravity duality'.

### 3.3.1. A weak/strong coupling duality

Gauge-gravity duality is a case of weak/strong coupling duality similar to the ones described above, since when the coupling of the bulk gravity theory is weak (viz. far away from the centre of the bulk) the gauge theory is strongly coupled. This can be seen as follows: both theories have two parameters in terms of which one can do a perturbative expansion (recall section 3.1). In the gravity theory, we have Newton's constant, $G_N$, and the radius of curvature of the AdS space, $\ell$ (Newton's constant is proportional to the string length, $\alpha'$).[30] On the other hand, in the gauge theory we have the coupling constant, $g$ (which determines the strength of the interactions), and the rank of the gauge group, $N$ (the number of colours in the theory; for Quantum Chromodynamics this would be $N = 3$). These parameters are related between the two theories as follows (see De Haro et al. (2017: section 4.1.2)):

$$\frac{G_N}{\ell^3} = \frac{\pi}{2N^2} \tag{4}$$

$$\frac{\ell^4}{\alpha'^2} = g^2 N, \tag{5}$$

where the parameters on the left-hand side are those of the gravitational theory, and the parameters on the right-hand side are those of the gauge theory.

---

[27] For more detail on this case and the successive extension of the idea of generalized EM duality to the context of string theory, see Castellani (2017, section 2).
[28] The space is actually only required to be *asymptotically, locally* AdS, rather than pure AdS.
[29] The QFT does not actually need to need to be exactly scale invariant. It is sufficient that it has a conformal fixed point.

Gravity is weak if Newton's constant, $G_N$, is much smaller than the radius of curvature of the AdS space, $\ell$, so that $\frac{G_N}{\ell^3} \ll 1$. Eq. (4) above then implies that the number of colours has to be large, i.e. $N \gg 1$. Also, quantum corrections will be suppressed if the radius of curvature of the space is much larger than the string length, $\frac{\ell^4}{\alpha'^2} \gg 1$ (so that the effects of the finite string length cannot be seen, and we basically deal with a point-particle theory), so that Eq. (5) gives $g^2 N \gg 1$. Now it was argued by 't Hooft (1974) that, in a gauge theory with $N$ colours, the natural coupling constant is not $g$, but rather the combination $g^2 N$. In other words, when $g^2 N \gg 1$ perturbation theory in the gauge theory breaks down, because the theory is strongly coupled. This is why the weak-gravity, semi-classical regime of the string theory corresponds to a strongly coupled gauge theory. The converse of this statement is of course also true: if the gauge theory is taken to be weakly coupled, so that $g^2 N \ll 1$, then the semi-classical (gravity) approximation to the string theory cannot be trusted, because an infinite tower of string corrections will give non-zero contributions.

Thus, the way gauge-gravity duality is a weak/coupling duality is similar, to some extent, to the example of electric-magnetic duality, in that the coupling constants of the two theories are inversely proportional to each other. However, gauge-gravity duality brings in a new element: the coupling constants do not have fixed values on the two sides, but can vary according to the details of a specific physical situation.[31] Let us just note, here, that the gauge theory coupling is a function of momentum, $g(k)$, while the string theory coupling is a function of the position in the AdS space. The region in which the gravity approximation is valid (i.e. the region where the gravity coupling is weak) is the region far away from the centre of the AdS space (if there is e.g. a black hole at the centre of AdS, the curvature will be strong). So, weak coupling requires large distances, far away from the centre, on the gravity side. But, as we saw above, weak gravity coupling is dual to strong coupling in the gauge theory, which happens at high momenta (and hence high energies).

In short: large distances (weak coupling) in the gravity theory correspond to high energies, i.e. small distances (strong coupling) in the dual gauge theory, and vice-versa. Thus, this is analogous to, although distinct from, the Fourier transformation example discussed in section 2.1.

One new element of gauge-gravity duality is the fact that the motion from the boundary towards the centre of the space, in the gravity theory, increases the gravitational coupling of the theory because the curvature radius increases. The dual of this inward motion, in the gauge theory, is motion from the UV towards the IR, i.e. towards low energies. This 'motion', which is interpreted in terms of spatial variation in the gravity theory but in terms of energy variation in the gauge theory, is called the 'renormalization group flow' of a gauge theory.

---

[30] These parameters can be written alternatively in string theory language, in terms of the string length (squared), $\alpha'$, and the string coupling, $g_s$. The string length determines how a string differs from a point particle, and the string coupling determines the perturbative expansion of the string theory. The expressions given in Eqs. (4) and (5) are for a five-dimensional gravity theory, which is the original example considered by Maldacena (1997).
[31] For more details, see Dieks et al. (2015, p. 207).

### 3.3.2. Fundamentality and emergence in gauge-gravity duality

Let us now take stock of what we said so far. Recall, from section 2.1, that the top and bottom theories are distinguished by the value of a parameter. In the gauge theory, the parameter is the momentum scale, which is dual to the radial direction in the gravity theory. Thus, we naturally get the diagram in Figure 1, where the vertical direction corresponds to the 'motion' discussed above (the RG flow), while the horizontal direction is the duality map.

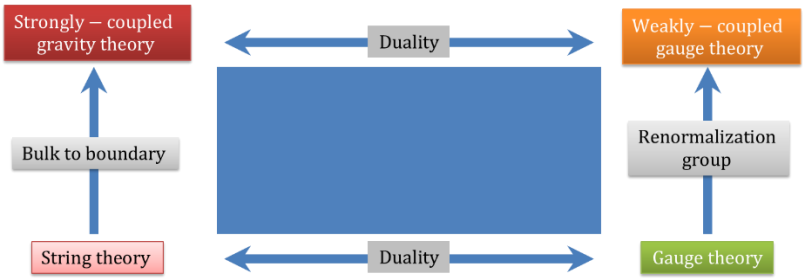

Figure 1. Duality relations vs. renormalization group flow.

In Figure 1, we have a horizontal relation between two theories which are dual, where the duality inverts the values of the couplings, according to Eqs. (4) − (5). But, in addition, we also have a vertical direction, which corresponds to the 'motion' in coupling space: radial motion on the gravity side, vs. the renormalization group flow on the gauge theory side. At each value of the coupling in the vertical direction, we have a pair of theories that are dual to each other.

Thus, as illustrated in Figure 1, we have two directions in which to consider the emergence and fundamentality issues: viz. the vertical and the horizontal directions. Indeed, the theories on Figure 1's bottom are the exact, non-perturbative theories, and (under the assumption of an exact gauge-gravity duality) they are dual to each other. These are the fundamental theories (assuming fundamentality in the sense specified in section 1.2), and the top-row theories are effective theories, and therefore less fundamental than these. Thus fundamentality increases in the downward vertical direction, but not in the horizontal direction. Emergence in the vertical direction is discussed in detail in Dieks et al. (2015: p. 210) and De Haro (2015: pp. 118-120), with the conclusion that there *is* ontological emergence in this direction. For example, on the gravity side we have Einstein's theory of general relativity with specific matter fields emerging in the low-energy limit of the underlying string theory. So far, we have the ordinary situation for effective field theories—that is, option (i).

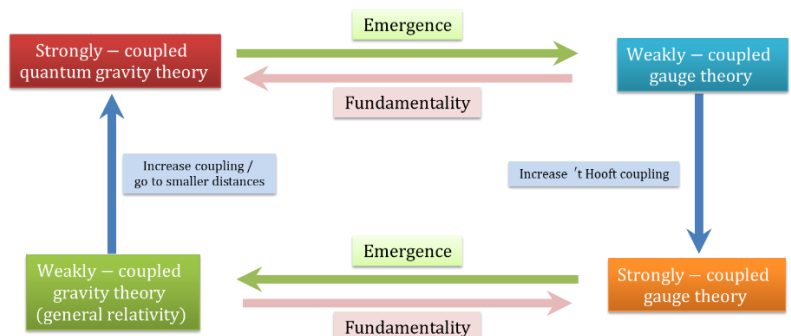

Figure 2. Emergence and fundamentality.

To get option (iii), we need to change the picture slightly, so that the theories on the bottom row are the weakly coupled string theory (i.e. the semi-classical gravity theory, general relativity) and the strongly coupled gauge theory which is dual to it (see Figure 2). In that case, the duality relation relates a weakly coupled theory to a strongly coupled theory. For this duality, there is a slight difficulty in identifying what is composite/solitonic and what is elementary, because we lack good descriptions of the strongly coupled theories. Nevertheless, we can still identify the weakly coupled theory as the more fundamental description, in the epistemic sense discussed in sections 1 and 2.2, that it is the description in which calculations can be done reliably: and we can identify its strongly coupled dual as the less fundamental one, since that description is out of control, when the coupling is weak in the other theory.[32]

Let us now examine epistemic emergence in the horizontal direction, i.e. along the duality. The question is whether the duality considered here can give rise to emergence. Notice that regarding ontological emergence, Dieks et al. (2015: p. 209) and De Haro (2015: p. 118) concluded that there cannot be any, because the two theories are exactly dual (i.e. equivalent), and therefore there can be no novelty, and so no emergence of one theory from the other.[33]

But this verdict can be modified when we consider emergence in the epistemic sense, i.e. as irreducibility or novelty of description, and fundamentality not as a property of the full theory, but as a property of the particular theoretical description. In this sense, one can indeed say that the weakly-coupled gravity description emerges from the strongly coupled gauge theory. Namely, imagine that one is working within the strongly coupled gauge theory, so that one is in a highly quantum regime where the usual perturbative description breaks down, and is unable to make any predictions. And assume that one then stumbles upon the duality, which points to a useful change of variables (in this case, an exceedingly complicated—presumably infinitely long—change of variables!), that allows one to reformulate the theory as a semi-classical gravity theory. That is, thanks to the duality it becomes possible to identify a set of variables that are more fundamental in this regime since, unlike the old variables, they give a good description, and can be the starting point for arriving at the gravity theory. The interpretation of these variables is not at all in terms of quantum fields, but in terms of semi-classical gravity. In this case, the gravity theory (and the objects within it) is indeed epistemically emergent.

Note that emergence here involves not *just* the change of variables (which is the mechanism (1) for emergence in section 2.2), but also a weak/coupling duality (mechanism (2)), and that this change of variables involves coarse-graining (mechanism (3)).[34]

## 4. Discussion and conclusions

The initial question motivating our contribution was the way in which the notions of fundamentality, emergence and duality can be intertwined, and how this connection can shed new light on fundamentality. More precisely, the novel feature is based on duality: how dualities are applied in contemporary physics, in particular the weak/string coupling duality, and the implications of this kind of duality for the philosophical discussion of fundamentality. Our starting point was to consider the three ways, (i) to (iii), of considering the relation between fundamentality and emergence.

---

[32] It is also very likely that there is a story about what is elementary and what is component in each description, like in the cases discussed in section 3.2: but we will set this issue aside.

[33] This verdict is subject to a specific interpretation, namely a so-called *internal interpretation.* Notice that from the mere presence of a duality, which is a formal relation, one cannot make a verdict about ontological emergence. To that end, one needs to consider the interpretation of the two theories, which in Dieks et al. (2014) and De Haro (2015) was done for internal interpretations.

[34] Emergence through a change of variables has also been considered recently by Franklin and Knox (2018), who define emergence in terms of 'novel explanation'. According to Franklin and Knox, changes of variables can give novel explanations of phenomena, because using the appropriate variables allows one to abstract away from irrelevant details and capture the salient explanatory elements. Although epistemic emergence is for us a matter of lack of derivability and therefore of descriptive, rather than explanatory, novelty, the role that changes of variables play in Franklin and Knox's account is similar to ours (another important difference is that, because of the weak/strong coupling dualities in this paper, our changes of variables often contain infinitely many terms, as well as coarse-graining).

While (i) and (ii) are commonly discussed in the literature on fundamentality in physics, duality suggests (iii) as a new way to construe the relation between emergence and fundamentality. We illustrated this in the two cases taken from quantum field theory (sine-Gordon/massive Thirring duality, and generalized EM duality) and in the case of gauge-gravity duality. The conclusion was similar for all these cases: the physical effective entities and the theories, as they are described, *can* play a more or less fundamental role, depending on the description chosen. And as such, there is emergence of a more fundamental description (because more elementary, in the sense specified in sections 2.2 and 2.3) out of a strongly coupled description. Notice that the direction of emergence is here *opposite* to that of fundamentality. Ordinarily, it is the non-elementary effective entities that emerge at lower energy.

Note that this 'inversion of the direction' is possible because our notions of fundamentality and emergence are epistemic. What is fundamental in this sense is not fixed once and for all by the ontology, but depends upon the description. Thus a more fundamental description can emerge within a strongly coupled theory. And this is made possible by the fact that quantum field theories can have more than one classical limit. In general, we expect that each classical limit will have its own emergent effective entities.

Let us underline that the kind of weak epistemic emergence found here is tightly connected with the notion of perturbation theory discussed in section 3.1. This is also the reason why we found no emergence in the cases of classical electromagnetism (section 3.2.1) and Dirac's quantization of it (section 3.2.2), where there are no perturbative expansions or perturbative duality, but only an exact one. It is only in the more sophisticated cases of quantum field theories and string theories that we get sufficient complexity to allow this kind of emergence.

As a final remark, let us recall that, in the example of gauge/gravity dualities, a spatial direction and the curvature of space emerge, together with the gravitational force, from a quantum field theory. This is of course an intriguing form of emergence of (at least one dimension of) space, along with new topological and geometric structures of the spacetime. We hope that regarding this as a case of epistemic emergence can cast light on the emergence of spacetime in theories of quantum gravity more generally: namely, as the emergence of spacetime from a reformulation of a non-spatiotemporal theory.

**Acknowledgements**

We thank Jeremy Butterfield and three referees for their comments. We thank the Vossius Center for History of Humanities and Sciences for its support, as well as the members of the Amsterdam Philosophy and History of Gravity group, for a discussion of the manuscript.